\documentclass[12pt,a4paper]{article}
\usepackage{graphics,graphicx}
\usepackage{amssymb,epsfig,amsmath,euscript,array}
\usepackage{cite}
\makeatletter
\@addtoreset{equation}{section}
\makeatother
\renewcommand{\theequation}{\thesection.\arabic{equation}}

\newcommand{\startappendix}{
\setcounter{section}{0}
\renewcommand{\thesection}{\Alph{section}}
\renewcommand{\theequation}{\Alph{section}.\arabic{equation}}}

\newcommand{\Appendix}[1]{
\refstepcounter{section}
\begin{flushleft}
{\Large\bf Appendix \thesection: #1}
\end{flushleft}}

\newcounter{multieqs}

\newenvironment{pretty}{}{}

\newcommand{\be}{\begin{equation}}
\newcommand{\ee}{\end{equation}}

\newcommand{\bm}[1]{\mbox{\boldmath $#1$}}

\newcommand{\kslash}{k \!\!\! / }

\newcommand{\lslash}{l \!\! / }
\newcommand{\Pslash}{P \!\!\!\! / }

\newcommand{\islash}{i \!\!\! / }
\newcommand{\jslash}{j \!\!\! / }
\newcommand{\aslash}{a \!\!\! / }
\newcommand{\bslash}{{b \hspace{-6pt} \slash} }

\newcommand{\onslash}{1 \!\!\! / }
\newcommand{\twslash}{2 \!\!\!/ }
\newcommand{\thslash}{3 \!\!\!/ }
\newcommand{\foslash}{4 \!\!\! / }
\newcommand{\fislash}{5 \!\!\! / }

\newcommand{\mslash}{m \!\!\! / }

\def\bd{\begin{document}}
\def\ed{\end{document}}
\def\nn{\nonumber}
\def\bea{\begin{eqnarray}}
\def\eea{\end{eqnarray}}

\def\ab{(ijab)}
\def\ba{(ijba)}
\def\ijab{{\tr}_{+}(\islash\, \jslash\, \aslash \, \bslash)}
\def\ijba{{\tr}_{+}(\islash\, \jslash\, \bslash \, \aslash)}
\def\ijaP{{\tr}_{+}(\islash\, \jslash\, \aslash \, \Pslash)}
\def\ijPLa{{\tr}_{+}(\islash\, \jslash\, \Pslash_L \, \aslash)}
\def\ijaPL{{\tr}_{+}(\islash\, \jslash\, \aslash \, \Pslash_L)}
\def\ijPLza{{\tr}_{+}(\islash\, \jslash\, \Pslash_{L;z} \, \aslash)}
\def\ijaPLz{{\tr}_{+}(\islash\, \jslash\, \aslash \, \Pslash_{L;z})}
\def\ijPa{{\tr}_{+}(\islash\, \jslash\, \Pslash \, \aslash)}
\def\iaPb{{\tr}_{+}(\islash\, \aslash\, \Pslash \, \bslash)}
\def\ibPa{{\tr}_{+}(\islash\, \bslash\, \Pslash \, \aslash)}
\def\ijPmu{{\tr}_{+}(\islash\, \jslash\, \Pslash \, \mu)}
\def\ibmuP{{\tr}_{+}(\islash\, \bslash\, \mu \, \Pslash)}
\def\ibmua{{\tr}_{+}(\islash\, \bslash\, \mu \, \aslash)}
\def\iamub{{\tr}_{+}(\islash\, \aslash\, \mu \, \bslash)}
\def\jaPb{{\tr}_{+}(\jslash\, \aslash\, \Pslash \, \bslash)}
\def\ijmuP{{\tr}_{+}(\islash\, \jslash\, \mu \, \Pslash)}
\def\ijmum{{\tr}_{+}(\islash\, \jslash\, \mu \, \mslash)}
\def\ijmmu{{\tr}_{+}(\islash\, \jslash\, \mslash \, \mu)}
\def\ijmP{{\tr}_{+}(\islash\, \jslash\, \mslash \, \Pslash)}
\def\iabP{{\tr}_{+}(\islash\, \aslash\, \bslash \, \Pslash)}
\def\ijbP{{\tr}_{+}(\islash\, \jslash\, \bslash \, \Pslash)}
\def\jbPa{{\tr}_{+}(\jslash\, \bslash\, \Pslash \, \aslash)}
\def\ijPb{{\tr}_{+}(\islash\, \jslash\, \Pslash \, \bslash)}
\def\jbmua{{\tr}_{+}(\jslash\, \bslash\, \mu \, \aslash)}

\def\loablt{ {\tr}_{+}(\lslash_1\, \aslash \, \bslash\, \lslash_2)}

\def\ijlolt{{\tr}_{+}(\islash\, \jslash\, \lslash_1 \, \lslash_2)}
\def\ijltlo{{\tr}_{+}(\islash\, \jslash\, \lslash_2 \, \lslash_1)}
\def\ibloa{{\tr}_{+}(\islash\, \bslash\, \lslash_1 \, \aslash)}
\def\jaltb{{\tr}_{+}(\jslash\, \aslash\, \lslash_2 \, \bslash)}
\def\ialtb{{\tr}_{+}(\islash\, \aslash\, \lslash_2 \, \bslash)}
\def\bltloa{{\tr}_{+}(\bslash\, \lslash_2\, \lslash_1 \, \aslash)}
\def\jbloa{{\tr}_{+}(\jslash\, \bslash\, \lslash_1 \, \aslash)}
\def\ibPb{{\tr}_{+}(\islash\, \bslash\, \Pslash \, \bslash)}
\def\ijltb{{\tr}_{+}(\islash\, \jslash\, \lslash_2 \, \bslash)}

\def\ijloa{{\tr}_{+}(\islash\, \jslash\,  \lslash_1 \, \aslash)}
\def\ijblt{{\tr}_{+}(\islash\, \jslash\,  \bslash \, \lslash_2)}

\def\jakb{{\tr}_{+}(\jslash\, \aslash\, \kslash \, \bslash)}
\def\iakb{{\tr}_{+}(\islash\, \aslash\, \kslash \, \bslash)}

\def\tofo{{\tr}_{+}(\onslash\, \thslash\, \twslash \, \foslash)}
\def\foto{{\tr}_{+}(\onslash\, \thslash\, \foslash \, \twslash)}
\def\tofi{{\tr}_{+}(\onslash\, \thslash\, \twslash \, \fislash)}
\def\fito{{\tr}_{+}(\onslash\, \thslash\, \fislash \, \twslash)}

\def\lrangle#1#2{\langle #1\,#2\rangle}

\def\Li{{$\rm Li}_2$}

\let\bm=\bibitem
\let\la=\label

\def\npb#1#2#3{Nucl. Phys. {\bf{B#1}} #3 (#2)}
\def\plb#1#2#3{Phys. Lett. {\bf{#1B}} #3 (#2)}
\def\prl#1#2#3{Phys. Rev. Lett. {\bf{#1}} #3 (#2)}
\def\prd#1#2#3{Phys. Rev. {D \bf{#1}} #3 (#2)}
\def\cmp#1#2#3{Comm. Math. Phys. {\bf{#1}} #3 (#2)}
\def\cqg#1#2#3{Class. Quantum Grav. {\bf{#1}} #3 (#2)}
\def\nppsa#1#2#3{Nucl. Phys. B (Proc. Suppl.) {\bf{#1A}}#3 (#2)}
\def\ap#1#2#3{Ann. of Phys. {\bf{#1}} #3 (#2)}
\def\ijmp#1#2#3{Int. J. Mod. Phys. {\bf{A#1}} #3 (#2)}
\def\rmp#1#2#3{Rev. Mod. Phys. {\bf{#1}} #3 (#2)}
\def\mpla#1#2#3{Mod. Phys. Lett. {\bf A#1} #3 (#2)}
\def\jhep#1#2#3{J. High Energy Phys. {\bf #1} #3 (#2)}
\def\atmp#1#2#3{Adv. Theor. Math. Phys. {\bf #1} #3 (#2)}
\newcommand{\EQ}[1]{\begin{equation} #1 \end{equation}}
\newcommand{\AL}[1]{\begin{subequations}\begin{align} #1 \end{align}\end{subequations}}
\newcommand{\SP}[1]{\begin{equation}\begin{split} #1 \end{split}\end{equation}}
\newcommand{\ALAT}[2]{\begin{subequations}\begin{alignat}{#1} #2 \end{alignat}
                        \end{subequations}}
\def\beqa{\begin{eqnarray}}
\def\eeqa{\end{eqnarray}}
\def\beq{\begin{equation}}
\def\eeq{\end{equation}}
\def\sst{\scriptscriptstyle}
\def\thetabar{\bar\theta}
\def\Tr{{\rm Tr}}
\def\one{\mbox{1 \kern-.59em {\rm l}}}
 \def\Nh{\hat{N}}

\def\a{\alpha}      \def\da{{\dot\alpha}}
\def\b{\beta}       \def\db{{\dot\beta}}
\def\c{\gamma}  \def\G{\Gamma}  \def\cdt{\dot\gamma}
\def\d{\delta}  \def\D{\Delta}  \def\ddt{\dot\delta}
\def\e{\epsilon}        \def\vare{\varepsilon}
\def\f{\phi}    \def\F{\Phi}    \def\vvf{\f}
\def\h{\eta}
\def\k{\kappa}
\def\l{\lambda} \def\L{\Lambda}
\def\m{\mu} \def\n{\nu}
\def\o{\omega}
\def\p{\pi} \def\P{\Pi}
\def\r{\rho}
\def\s{\sigma}  \def\S{\Sigma}
\def\t{\tau}
\def\th{\theta} \def\Th{\Theta} \def\vth{\vartheta}
\def\X{\Xeta}
\def\z{\zeta}
\def\cA{{\cal A}} \def\cB{{\cal B}} \def\cC{{\cal C}}
\def\cD{{\cal D}} \def\cE{{\cal E}} \def\cF{{\cal F}}
\def\cG{{\cal G}} \def\cH{{\cal H}} \def\cI{{\cal I}}
\def\cJ{{\cal J}} \def\cK{{\cal K}} \def\cL{{\cal L}}
\def\cM{{\cal M}} \def\cN{{\cal N}} \def\cO{{\cal O}}
\def\cP{{\cal P}} \def\cQ{{\cal Q}} \def\cR{{\cal R}}
\def\cS{{\cal S}} \def\cT{{\cal T}} \def\cU{{\cal U}}
\def\cV{{\cal V}} \def\cW{{\cal W}} \def\cX{{\cal X}}
\def\cY{{\cal Y}} \def\cZ{{\cal Z}}
\def\ua{\underline{\alpha}}
\def\ub{\underline{\phantom{\alpha}}\!\!\!\beta}
\def\uc{\underline{\phantom{\alpha}}\!\!\!\gamma}
\def\um{\underline{\mu}}
\def\ud{\underline\delta}
\def\ue{\underline\epsilon}
\def\una{\underline a}\def\unA{\underline A}
\def\unb{\underline b}\def\unB{\underline B}
\def\unc{\underline c}\def\unC{\underline C}
\def\und{\underline d}\def\unD{\underline D}
\def\une{\underline e}\def\unE{\underline E}
\def\unf{\underline{\phantom{e}}\!\!\!\! f}\def\unF{\underline F}
\def\unm{\underline m}\def\unM{\underline M}
\def\unn{\underline n}\def\unN{\underline N}
\def\unp{\underline{\phantom{a}}\!\!\! p}\def\unP{\underline P}
\def\unq{\underline{\phantom{a}}\!\!\! q}
\def\unQ{\underline{\phantom{A}}\!\!\!\! Q}
\def\unH{\underline{H}}
\def\As {{A \hspace{-6.4pt} \slash}\;}
\def\bs {{b \hspace{-6.4pt} \slash}\;}
\def\Ds {{D \hspace{-6.4pt} \slash}\;}
\def\ds {{\del \hspace{-6.4pt} \slash}\;}
\def\ss {{\s \hspace{-6.4pt} \slash}\;}
\def\ks {{ k \hspace{-6.4pt} \slash}\;}
\def\ps {{p \hspace{-6.4pt} \slash}\;}
\def\pas {{{p_1} \hspace{-6.4pt} \slash}\;}
\def\pbs {{{p_2} \hspace{-6.4pt} \slash}\;}
\def\Ps {{P \hspace{-6.4pt} \slash}\;}
\def\Qs {{Q \hspace{-6.4pt} \slash}\;}
\def\Fh{\hat{F}}
\def\Vh{\hat{V}}
\def\Xh{\hat{X}}
\def\ah{\hat{a}}
\def\xh{\hat{x}}
\def\yh{\hat{y}}
\def\ph{\hat{p}}
\def\xih{\hat{\xi}}
\def\psit{\tilde{\psi}}
\def\Psit{\tilde{\Psi}}
\def\tht{\tilde{\th}}
\def\lt{\tilde{\lambda}}
\def\llt{\tilde{l}}
\def\At{\tilde{A}}
\def\Qt{\tilde{Q}}
\def\Rt{\tilde{R}}
\def\Nt{\tilde{N}}

\def\at{\tilde{a}}
\def\st{\tilde{s}}
\def\ft{\tilde{f}}
\def\pt{\tilde{p}}
\def\qt{\tilde{q}}
\def\vt{\tilde{v}}
\def\nt{\tilde{n}}
\def\delb{\bar{\partial}}
\def\bz{\bar{z}}
\def\bD{\bar{D}}
\def\bB{\bar{B}}
\def\bk{{\bf k}}
\def\bl{{\bf l}}
\def\bp{{\bf p}}
\def\bq{{\bf q}}
\def\br{{\bf r}}
\def\bx{{\bf x}}
\def\by{{\bf y}}
\def\bR{{\bf R}}
\def\bV{{\bf V}}
\def\d{\delta}\def\D{\Delta}\def\ddt{\dot\delta}
\def\pa{\partial} \def\del{\partial}
\def\xx{\times}
\def\uno{\mbox{1 \kern-.59em {\rm l}}}
\def\trp{^{\top}}
\def\inv{^{-1}}
\def\dag{{^{\dagger}}}
\def\pr{^{\prime}}
\def\lan{\langle}
\def\ran{\rangle}
\def\rar{\rightarrow}
\def\lar{\leftarrow}
\def\lrar{\leftrightarrow}
\newcommand{\0}{\,\!}      %this is just NOTHING!
\def\one{1\!\!1\,\,}
\def\im{\imath}
\def\jm{\jmath}
\newcommand{\tr}{\mbox{tr}}
\newcommand{\slsh}[1]{/ \!\!\!\! #1}
\def\vac{|0\rangle}
\def\lvac{\langle 0|}
\def\hlf{\frac{1}{2}}
\def\ove#1{\frac{1}{#1}}
\def\Box{\square}
\def\ZZ{\mathbb{Z}}
\def\CC#1{({\bf #1})}
\def\bcomment#1{}
\def\bfhat#1{{\bf \hat{#1}}}
\def\VEV#1{\left\langle #1\right\rangle}
\newcommand{\ex}[1]{{\rm e}^{#1}} \def\ii{{\rm i}}
\def\rr{{\rm r}} \def\rs{{\rm s}}\def\rv{{\rm v}}
\def\ri{{\rm i}}\def\rj{{\rm j}}
\newcommand{\lrbrk}[1]{\left(#1\right)}
\newcommand{\sfrac}[2]{{\textstyle\frac{#1}{#2}}}

\def\Li{{\rm Li}_2}

\font\mybb=msbm10 at 12pt
\def\bb#1{\hbox{\mybb#1}}

\font\myBB=msbm10 at 18pt
\def\BB#1{\hbox{\myBB#1}}

\begin{document}

\begin{flushright}
hep-th/0502146 \\
QMUL-PH-05-02
\end{flushright}

\vspace{20pt}

\begin{center}

{\Large \bf A Recursion Relation for Gravity Amplitudes}
%\\
%\vspace{0.3cm}
%{\Large \bf  for Gravity Amplitudes}
\vspace{12pt}
\vspace{33pt}
%%%%%%%%%%%%%%%%%%%%%%%
%\\
%{\Large \bf  OR: The Analytic S-Matrix, Once Again!}
%\vspace{12pt}

%%%%%%%%%%%%%%%%%%%%%%%
{\bf James Bedford, Andreas  Brandhuber, Bill  Spence and Gabriele  Travaglini}
\begin{pretty}\footnote{
{\sffamily \{\tt j.a.p.bedford, a.brandhuber, w.j.spence,
g.travaglini\}@qmul.ac.uk }}\end{pretty}

{\em Department of Physics\\
Queen Mary, University of
London\\
Mile End Road, London, E1 4NS\\
United Kingdom
 }

\vspace{40pt} {\bf Abstract}

\end{center}

% ABSTRACT goes here

\noindent 
Britto, Cachazo and Feng have recently derived a 
recursion relation for tree-level scattering amplitudes 
in Yang-Mills. This relation has a bilinear structure 
inherited from factorisation on multi-particle poles 
of the scattering amplitudes -- a rather generic feature 
of field theory. 
Motivated by this, we propose a new recursion relation 
for scattering amplitudes of gravitons at tree level.
Using this, we derive a new general formula for the 
MHV tree-level scattering amplitude for 
$n$ gravitons.
Finally, we comment on the existence of
recursion relations in general field theories.

\vspace{0.5cm}

\setcounter{page}{0}
\thispagestyle{empty}
\newpage

%%%%%%%%%%%%%%%%%%%%%%%%%%%%%%%%%%%%%%%%%%%%%%%%%%%%%%%%%%%%%%%%%
%%%%%%%%%%%%%%%%%%%%%%%%%%%%%%%%%%%%%%%%%%%%%%%%%%%%%%%%%%%%%%%%%

\section{Introduction}
\setcounter{footnote}{0}

Much progress has been made in the past year 
in understanding  the structure and 
practical calculation of scattering amplitudes 
in four-dimensional Yang-Mills theories \cite{oxford}. 
This has been prompted by the conjecture that 
twistor string theory provides a dual description of 
%tree-level S-matrix elements in 
weakly-coupled gauge theory \cite{witten}. 
% with results from earlier unitarity-based methods
%\cite{Bern:1994cg,Bern:zx}

Outstanding progress in the efficient calculation 
of scattering amplitudes was achieved 
by Cachazo, Svr\v{c}ek and Witten 
(CSW) \cite{csw}, 
who proposed a radical new method for computing 
tree-level amplitudes in Yang-Mills using 
MHV amplitudes, appropriately continued off shell, as vertices. 
This procedure was later extended to one loop  in 
\cite{bst}, and used to re-derive one-loop MHV scattering 
amplitudes in $\cN\!=\! 4$ super Yang-Mills. 

%Subsequently, the same procedure was applied to 
%$\cN\!=\! 1$ super Yang-Mills \cite{quig,bbst1}, where 
%the corresponding MHV amplitudes where re-calculated, and 
%finally to pure Yang-Mills in \cite{bbst2}, 
%where the cut-constructible part of the infinite sequence 
%of MHV amplitudes was found, generalising earlier results 
%of \cite{bdk9302280}.  

Further progress in the calculation 
of one-loop amplitudes was achieved
using the cut-constructibility approach
\cite{Bern:1994cg,Bern:zx}
in \cite{new}, 
CSW diagrams \cite{quig,bbst1,higgs,bbst2},
the holomorphic anomaly  \cite{csw3,freddy,freddy2,BBDD}
and generalized cuts \cite{Britto,rsv04,bdk04}. 
An interesting spin-off of the latter approach
was that the new results for loop amplitudes 
could be used to find new
representations of tree-level amplitudes \cite{bdk04}. 
This is a direct consequence of
the structure of infrared singularities of one-loop scattering 
amplitudes in gauge
theory. Inspired by these insights,  
a new recursion formula for tree-level scattering
amplitudes was proposed in \cite{bcf}, 
which is quadratic in the amplitudes and leads
to very compact formulae. 
Recently, this proposal was proved in  \cite{bcfw} using
analyticity and factorization properties 
of gauge theory amplitudes. 
Recurrence relations were also derived in \cite{Bern:2005hs}
to determine the rational part of one-loop amplitudes in 
QCD. 

%
%An interesting recent development has been the proposal of
%a new recursion relation for scattering amplitudes
%in Yang-Mills at tree level \cite{bcf}, 
%which was subsequently proved in
%\cite{bcfw}. A new class of representations of tree amplitudes had
%been found in \cite{bdk04}; one of these suggested 
%the existence of new
%recursion relations for tree amplitudes \cite{rsv04}.

A key ingredient in the proof of the recursion relation
in \cite{bcfw} is the fact that
scattering amplitudes in Yang-Mills
factorise on multi-particle poles.
This is a fully non-perturbative statement
%(i.e.~its validity is not limited to tree-level scattering),
and a general property of field theory.%
\footnote{See, for example, chapter 10 of \cite{wein}.}
As such, it leads one to suspect that 
it should be possible to write down 
recursion relations for scattering amplitudes
in generic theories which admit 
a field-theoretical description.
In this paper we explore this idea, 
and propose a new recursion relation
for the tree-level scattering of gravitons.
The relation we prove is directly inspired by the
BCF/BCFW relation \cite{bcf,bcfw}, 
but takes into account the specific
features which arise when considering gravity amplitudes.

One of the appealing features of the recursion relation
of \cite{bcf} is that
it generates new formulae for amplitudes, 
which are often of a simpler form.
We will show that the gravity recursion relation
we propose also leads to
a new formula for the $n$-point MHV amplitude 
for gravity scattering, which agrees with 
an earlier formula derived by Berends, Giele and Kuijf
\cite{bgk}.  Guided by preliminary investigation
of next-to-MHV gravity amplitudes, 
we expect that these recursion relations
are also correct for more general tree-level gravity amplitudes.

The plan of the rest of the paper is as follows: 
In section 2, drawing inspiration from the 
BCFW proof \cite{bcfw} of the recursion relation 
for tree-level scattering amplitudes in Yang-Mills,
we derive a recursion relation 
for scattering amplitudes of gravitons at tree level. 
In section 3 we apply this recursion relation, and 
derive a new expression for the infinite sequence of 
MHV scattering amplitudes of gravitons. 
Finally, section 4 is devoted to some comments on 
recursion relations in other field theories. 

For other recent work 
on gravity amplitudes see
\cite{WuZhutwo,Giombi:2004ix,Nair:2005iv,Bern:2005bb}. 
For related work on gauge theory amplitudes, see
\cite{Zhu}--\!\!\cite{a}.

%%%%%%%%%%%%%%%%%

\section{The recursion relation in gravity}
In this section we closely follow the proof
of the recursion relation in Yang-Mills \cite{bcfw}, 
which we will extend to the case of gravity amplitudes.
As we shall see, the main new ingredient is that
gravity amplitudes depend on more kinematical invariants 
than the corresponding Yang-Mills amplitudes, 
namely those which are sums of non-cyclically adjacent momenta; 
hence, more multi-particle channels should be considered.

To derive a recursion relation for scattering amplitudes,
we start by introducing a one-parameter family of
scattering amplitudes, $\cM(z)$ \cite{bcfw},
where we choose $z$ in such a way that
$\cM(0)$ is the amplitude we wish to compute.
We  work in complexified Minkowski space and regard
$\cM(z)$ as a complex function of $z$ and the momenta.
One can then consider the contour integral 
\cite{Bern:2005hs}
\beq
\label{int}
\cC_{\infty} \ := \ 
{1\over 2\pi i} \oint\! dz \  { \cM(z) \over z }
\ ,
\eeq
where the integration is taken around the circle at
infinity in the complex $z$ plane.  
Assuming that $\cM (z)$ has only simple poles at $z=z_i$,
the integration gives
\beq
\cC_{\infty} \ = \
\cM(0) \, + \, \sum_i { [{\rm Res} \, \cM(z)]_{z=z_i}
\over z_i}
\ .
\eeq
In the important case of Yang-Mills amplitudes, 
$\cM(z) \to 0$ as $z \to \infty$,  
and hence $\cC_{\infty}=0$.

Notice that up to this point the definition of
the family of amplitudes  $\cM(z)$ has not been given 
-- we have not even specified the theory
whose scattering amplitudes we are computing.

There are some obvious requirements for $\cM(z)$. 
The main point is to define $\cM(z)$ in such a way that 
poles in $z$ correspond to multi-particle poles 
in the scattering amplitude  $\cM(0)$. 
If this occurs, then the corresponding residues 
% $c_k := \lim_{z \to z_k} (z-z_k) \cM(z)$
can be computed from factorisation properties 
of scattering amplitudes (see, for example, 
\cite{lance,wein}). 
In order to accomplish this, 
$\cM(z)$ was defined in \cite{bcf,bcfw}  
by shifting the momenta of two of the external particles
in the original scattering amplitude. 
For this procedure to make sense, we have to make sure 
that even with these shifts overall momentum conservation 
is preserved, and that all particle momenta remain on-shell.
We are thus led to define $\cM(z)$ as 
the scattering amplitude 
$\cM (p_1 , \ldots , p_k (z), \ldots , p_l(z), \ldots , p_n)$,
where the momenta of particles $k$ and $l$ are shifted to 
\beq
\label{shifts}
p_k (z) \ := \  p_k + z \eta \ ,  \qquad 
p_l(z)\  := \ p_l - z \eta
\ .
\eeq
Momentum conservation is then maintained.
As in \cite{bcf}, we can solve $p_k^2 (z) = p_l^2 (z) = 0$
by choosing $\eta = \l_l \lt_k$ (or  $\eta = \l_k \lt_l$),
which makes sense in complexified Minkowski space.
Equivalently, 
\beq
\label{shiftsspin}
\l_k (z) \ := \  \l_k + z \l_l \ ,  \qquad 
\lt_l (z)\  := \  \lt_l - z \lt_k
\ , 
\eeq
with $\l_l$ and $\lt_k$ unshifted.

More general families of scattering amplitudes
can also be defined, as pointed out in 
\cite{Bern:2005hs}.
For instance, one can single out
three particles $k$, $l$, $m$,
and define 
\beq
\label{shiftsgen}
p_k (z) \ := \  p_k +  z \eta_k \ ,  \qquad
p_l(z)\  := \ p_l + z \eta_l
\ ,  \qquad
p_m(z)\  := \ p_m + z \eta_m
\ ,
\eeq
where $\eta_k$, $\eta _l$ and $\eta_m$ are null and
$\eta_k + \eta_l + \eta_m =0$.
Imposing $p_k^2(z)= p_l^2(z)=p_m^2(z)=0$,
one finds the solution
\beq
\label{genshift}
\eta_k = - \a \l_k \lt_l - \b \l_k \lt_m
\ , \qquad
\eta_l = \a \l_k \lt_l \ , \qquad
\eta_m = \b \l_k \lt_m
\ ,
\eeq
for arbitrary $\a$ and $\b$.
This has been used in \cite{Bern:2005hs}.
In the following we will limit
ourselves to shifting only two momenta 
as in \cite{bcf} and \cite{bcfw}.

At tree level, scattering amplitudes in field theory
can only have simple poles in multi-particle channels; 
for $\cM (z)$, these generate  
poles in $z$  (unless the channel contains 
both particles $k$ and $l$, or none).  
Indeed, if $P(z)$ is a sum of momenta including $p_l(z)$
but not $p_k(z)$, then  $P^2 (z) = P^2 - 2 z (P \cdot \eta)$
vanishes at $z_P = P^2 / 2  (P\cdot \eta)$ \cite{bcfw}.
In Yang-Mills theory, one considers colour-ordered partial 
amplitudes, which have a fixed cyclic ordering of the 
external legs. This implies that a generic 
Yang-Mills partial amplitude can only depend 
on kinematical invariants made of 
sums of cyclically adjacent momenta.  
Hence, tree-level Yang-Mills amplitudes 
can only have poles  in kinematical channels 
made of cyclically adjacent sums of momenta.

For gravity amplitudes this is not the case, 
as there is no such notion of ordering for the external legs. 
Therefore, the multi-particle poles 
which produce poles in $z$ 
are those obtained by forming all possible combinations 
of momenta which include $p_k(z)$  but not $p_l(z)$. 
This is the only modification to the BCFW
recursion relation we need to make, in order 
to derive a gravity recursion relation.

For any such multi-particle channel  $P^2 (z)$, 
we have 
\beq
\cM (z) \to
\sum_{h}\,
{\cM_L^{h} (z_P)} \, {1 \over P^2 (z)}
\,
{\cM_R^{-h} (z_P)}
 \ ,
\eeq
as $P^2 (z) \to 0$ (or, equivalently, $z \to z_P$).
The sum is over the possible helicity assignments
on the two sides of the propagator
which connects the two lower-point tree-level 
amplitudes $\cM_L^{h}$ and $\cM_R^{-h}$.
It follows that
\beq
[{\rm Res} \, \cM (z) ]_{z=z_P} \ = \
 - \sum_{h}\,
\cM_L^{h} (z_P) \, {z_P\over P^2} \, \cM_R^{-h} (z_P)
\ ,
\eeq
so that finally
\beq
\cM(0)  \ = \
\cC_{\infty}\, + \,
\sum_{P, h} { \cM_L^{h} (z_P)  \, \cM_R^{-h} (z_P) \over P^2}
\ .
\eeq
The sum is over all possible decompositions 
of momenta such that
$p_k \in P$ but $p_l \notin P$.

If $\cC_{\infty} = 0$, then there is no boundary term
in the recursion relation, and
\beq
\label{rec}
\cM(0)  \ = \
\sum_{P, h} { \cM_L^{h} (z_P) \,  \cM_R^{-h} (z_P) \over P^2}
\ .
\eeq
In \cite{bcfw} it was shown that for Yang-Mills amplitudes
boundary terms $\cC_{\infty}^{\rm YM}$
always vanish. Two different proofs were presented, 
the first based on the use of CSW diagrams \cite{csw},
the second on Feynman diagrams.
For gravity, we still lack a description in terms
of MHV vertices, so we can only rely on Feynman diagrams.
This is also the case for other field theories we might be
interested in (such as $\lambda \phi^4$, for example).
As we have remarked, $\cC_{\infty} = 0$ if  $\cM (z)  \to 0$ 
as $z \to \infty$.
$\cM (z)$ is a scattering amplitude with shifted, $z$-dependent
external null momenta. One can then try to estimate
the behaviour of $\cM (z)$  for large  $z$
by using power counting
(different theories will of course give different results).
In $\l \phi^4$ the Feynman vertices are momentum independent
and  $\cC_{\infty} = 0$ (see the last section); 
in quantum gravity, however, vertices are quadratic in momenta,
and one cannot determine a priori whether or not a boundary term
is present.

From the previous discussion, it follows that
the behaviour of $\cM (z)$ as $z \to \infty$ is related to
the high-energy behaviour of the scattering amplitude 
(and hence to the renormalisability of the theory).
The ultraviolet behaviour of quantum gravity, 
however, is full of surprises 
(for a summary, see for example section (2.2) of
\cite{Bern:2002kj}).
We may therefore expect a more benign behaviour of 
$\cM (z)$ as $z \to \infty$. 
Specifically, in the next section we will focus on
the MHV scattering amplitudes of $n$ gravitons, which
have been computed by Berends, Giele and Kuijf (BGK) in
\cite{bgk}.
Performing the shifts \eqref{shifts} explicitly
in the BGK formula, 
one finds the surprising result%
\footnote{We have checked that 
$\cM(z) \sim \cO (1 / z^2 )$ as $z \to \infty$,  
analytically for $n\leq 7$ legs, and numerically 
for $n\leq 11$ legs.}
\beq
\lim_{z \to \infty} \cM_{\rm MHV}(z) \ = \ 0
\ .
\eeq
In more general amplitudes one can (at least in principle)
use the (field theory limit of the) 
KLT relations \cite{klt}, which connect 
tree-level gravity amplitudes to tree-level 
amplitudes in Yang-Mills, to estimate 
the large-$z$ behaviour of the scattering amplitude.%
\footnote{See the appendix for explicit examples
of KLT relations for four, five and six legs.}
As an example, we have considered  
the next-to-MHV gravity amplitude 
$\cM (1^- , 2^- , 3^- , 4^+ , 5^+ , 6^+)$, 
and performed the shifts as in \eqref{shiftsspin}, 
with $k=1$ and $l=2$. Similarly to the 
MHV case, we find that 
\beq
\lim_{z \to \infty} 
\cM (1^- , 2^- , 3^- , 4^+ , 5^+ , 6^+) (z)  \ = \ 0
\ . 
\eeq
It would be interesting to understand if this 
is true for generic gravity amplitudes. 
%In any event, at tree level one can perform more generic shifts
%such as \eqref{shiftsgen} 
%in order to eliminate possible boundary terms, 
%in the same spirit as \cite{Bern:2005hs}.

In the next section we will apply the recursion relation
\eqref{rec} to the case of MHV amplitudes in  gravity, 
and show that it does generate correct expressions 
for the amplitudes. As a bonus, we will derive a new 
closed form expression for the $n$-particle scattering amplitude.

Before moving to the explicit computations, we would like
to make some comments on how momentum shifts 
may be generated inside amplitudes.
To begin with, we should notice that there 
is an intriguing difference
between the MHV diagram method and the 
BCF recursion relation, namely the
fact that in the latter one sums only over 
a subset of channels -- those
where the reference legs $k$ and $l$ are 
on different sides with respect to the internal propagator.
It is clear that any derivation 
(rigorous or heuristic) of the recursion
relation from the MHV diagrams method, or vice versa, 
will have to address this point.  
In the following we limit ourselves to some formal
observations aimed at making a preliminary connection between 
the MHV diagram method and the 
shifts in the sub-amplitudes appearing in the 
recursion relation.

Let $\cA$ be a tree-level scattering amplitudes
of $n$ gluons in Yang-Mills, and
let us focus on a particular channel
$P_{ij} := p_i + \cdots + p_j$.
As $P_{ij}^2 \to 0$, the scattering amplitude factorises as
\beq
\cA \to \sum_{h} \cA_L^{h} (j+1, \ldots, i-1, P_{ij})  \,
{1\over P_{ij}^2}
\cA_R^{-h}(-P_{ij},  i, \ldots , j)
\ .
\eeq
Now imagine that we want to construct the full amplitude --
at $P^2_{ij} \neq  0$ -- from MHV diagrams.
The issue then arises of determining
the spinors associated with the non-null momentum
$P^2_{ij}$. A prescription equivalent 
to those of CSW consists of decomposing \cite{dk}%
\footnote{In the following we drop the subscript
$ij$ in $P_{ij}$.}
\beq
\label{dec}
P = \l \lt \, + \, z \, \eta
\ ,
\eeq
where $\eta$ is a reference null momentum.
$\l$ and $\lt$ are then the spinors associated with 
the momentum $P$.
The decomposition \eqref{dec}
played a central r\^{o}le in the calculation
of one-loop amplitudes from MHV diagrams
performed in \cite{bst,bbst1,bbst2}.
Attached to the scalar propagator with momentum
$P$ there will be two effective amplitudes $\cA_L$ and
$\cA_R$ computed from MHV diagrams. We will use $\l$ and
$\lt$ as spinors associated with $P$, and write
the contribution to the amplitude as
(schematically)
\beq
\cA_L (l, \ldots)\, {1 \over  P^2} \,  
\cA_R(-l , \ldots) \, \, 
\delta^{(4)} (P_L + P_R)
\ , 
\eeq 
where $P_L$ and $P_R$ are the sum of the external momenta
on the left and on the right of the diagrams, 
with $P= -P_L = P_R$,  and $l:=\l \lt$. 
We have also explicitly written a delta function
for momentum conservation, which we recast as
$ \int\!\! d^4P \, \delta^{(4)} (P_L + P)\,
\delta^{(4)} (P_R - P)$.

We wish to associate each delta function with 
the corresponding amplitude on the left or on the right
of the diagram. We also expand $P$ as in \eqref{dec}, so that
the result is
\beq
\int\!\! d^4P \, \delta^{(4)} (P_L + l + z \eta)\, \cA_L(l, \ldots)
\ {1\over P^2}\
\delta^{(4)} (-P_R + l + z \eta)\, \cA_R(-l, \ldots)
\ .
\eeq
Now, consider each sub-amplitude, for example the one on the left
which, together with its delta function, reads
\beq
\cA_L ( l , \ldots ) \, \delta^{(4)} (P_L + l + z \eta)
\ .
\eeq
The delta function can be interpreted as imposing the condition that
the sum of the external momenta is now $P_L + z \eta$, 
rather than $P_L$. 
Analogously, on the right this sum will be
$P_R - z \eta$. Overall momentum is trivially conserved,
but at the level of each sub-amplitude we have to shift
the sum of the external momenta by $\pm z \eta$.

One way to do this is to imagine that the extra momentum
is used entirely to shift the momentum of
a single particle on the left, say $p_k$,  by $z \eta$,
$p_k \to p_k(z) := p_k + z \eta$,
and the momentum of a single particle
on the right, $p_l$,  by $-z \eta$,
$p_l \to p_l(z) := p_l - z \eta$.
Imposing the requirement that the new momenta are still null leads to
$\eta = \l_k \lt_l$ or  $\eta = \l_l \lt_k$, as discussed before.
One could alternatively attribute the momentum shifts
to more than one particle, which would lead to the more generic
shifts \eqref{genshift}.

\section{Application to MHV  gravity amplitudes }
In the following we will compute the MHV scattering amplitude
$\cM(1^- , 2^- , 3^+ , \ldots , n^+)$
for  $n$ gravitons. 
We will choose the two negative helicity gravitons
$1^-$ and $ 2^-$
as reference legs.
This is a particularly convenient choice,
as it reduces the number of
terms arising in the recursion relation
to a minimum.
The shifts for the momenta of particles 1 and 2 are
\beq
p_1 \to p_1 +  z \l_2 \lt_1 \ ,
\qquad p_2 \to p_2 -  z \l_2 \lt_1
\  .
\eeq
In terms of spinors, the shifts are realised as
\beq
\label{shiftsonceagain}
\l_1 \to \hat{\l}_1 := \l_1 +  z \l_2 \ , \qquad
\lt_2 \to \hat{\lt}_2 := \lt_2 - z \lt_1
\  ,
\eeq
with $\l_2$ and $\lt_1$ unmodified.

\begin{figure}[ht]
\label{figure-vv}
\begin{center}
\scalebox{0.65}{\includegraphics{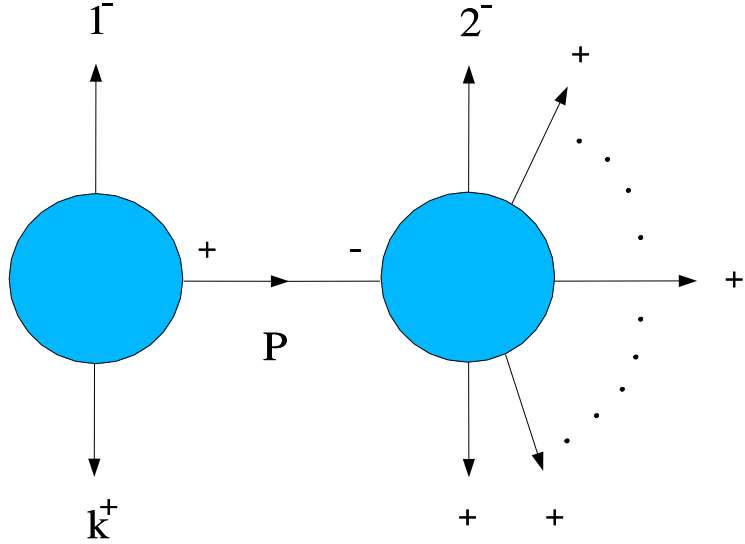}}
\end{center}
\caption{\it 
One of the  terms contributing to the recursion relation 
for the MHV  amplitude 
$\cM  (1^- , 2^- , 3^+ , \ldots , n^+)$. 
The gravity scattering amplitude on the right 
is symmetric under the exchange of gravitons
of the same helicity. In the recursion relation, we sum 
over all possible values of $k$, i.e.~$k=3, \ldots , n$. 
This amounts to summing over cyclical permutations of 
$(3, \ldots , n)$.
}
\end{figure}

\begin{figure}[ht]
\label{figure-vv2}
\begin{center}
\scalebox{0.65}{\includegraphics{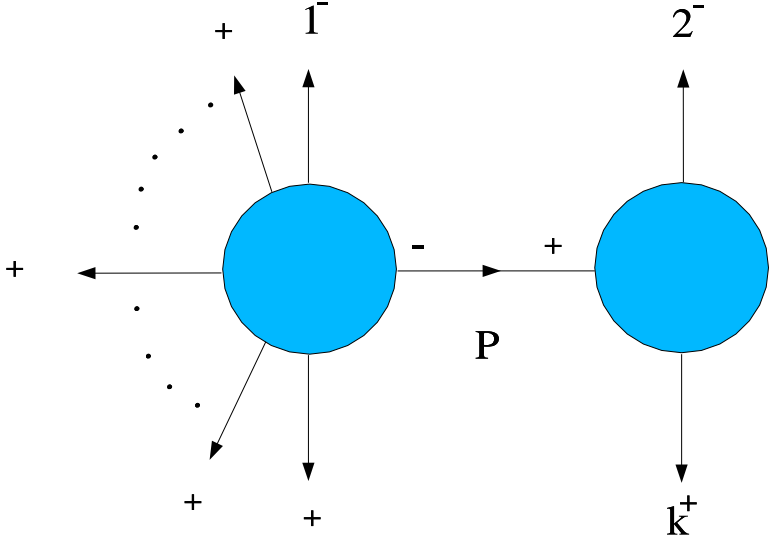}}
\end{center}
\caption{\it 
This class of diagrams also contributes 
to the recursion relation for the MHV amplitude 
$\cM  (1^- , 2^- , 3^+ , \ldots , n^+)$; however,  
each of these diagrams vanishes if 
the shifts \eqref{shiftsonceagain} 
are performed. 
}
\end{figure}

Let us consider the possible recursion diagrams
that can arise. There are only two possibilities,
corresponding to the two possible internal helicity assignments,
$(+-)$ and $(-+)$:
\begin{itemize}
\item[{\bf 1.}]
The amplitude on the left is googly $(++-)$,
whereas on the right there is an MHV gravity amplitude
with $n-1$ legs (see Figure 1).
\item[{\bf 2.}]
The amplitude on the right is googly, and the amplitude on
the left is MHV (see Figure 2).
\end{itemize}
We recall that a gravity amplitude is symmetric under 
the interchange of identical helicity gravitons; 
this implies that we have to sum $n-2$ diagrams for each 
of the configurations in Figures 1 and 2. 
Each diagram is then completely specified 
by choosing $k$, with $k=3, \ldots, n$.

However, it is easy to see that diagrams of the type {\bf 2.}
actually give a vanishing contribution.
Indeed, they are proportional to
\beq
[k\, \hat{P} ] \, = \,
{ [ k | \hat{P} | \hat{2} \ran \over \lan \hat{P} \, \hat{2} \ran}
\, = \,
{ [ k | P | 2 \ran \over \lan \hat{P} \, \hat{2} \ran} \, = \, 0
\ ,
\eeq
where the last equality follows from $P= p_k + p_2$.
Hence we will have to compute diagrams of type {\bf 1.} only.
We will do this in the following.

\subsection{Four, five and six graviton scattering}
To show explicitly how our recursion relation generates
amplitudes, we will now derive the $4,5$ and $6$ point MHV
scattering amplitudes.

We start  with the four point case. 
There are two diagrams to sum, one of which is represented
in Figure 3; the other is obtained by 
swapping the labels $4$ with $3$. 
\begin{figure}[ht]
\label{figure-vvv}
\begin{center}
\scalebox{0.65}{\includegraphics{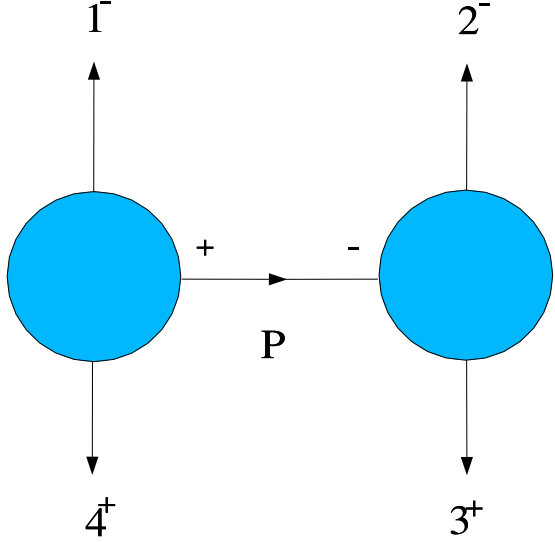}}
\end{center}
\caption{\it 
One of the two diagrams contributing to the recursion relation 
for the MHV  amplitude 
$\cM  (1^- , 2^- , 3^+ , 4^+)$.  
The other is obtained from this by cyclically permuting 
the labels $(3,4)$ -- i.e.~swapping $3$ with $4$.}
\end{figure}
For the diagram in Figure 3, we have 
\beq
\cM^{(4)} \ = \ \cM_{L} \, {1 \over P^2} \, \cM_R \ , 
\eeq
where the superscript denotes the label on the positive helicity 
leg in the trivalent vertex, 
\beqa
\cM_L & = & 
\left( 
{ [ \hat{P} \, 4]^3 \over [ 4 \, 1] [ 1 \, \hat{P} ] }
\right)^2 \ , 
\\ \nonumber \cr
\cM_R & = & 
\left( 
{ \lan  \hat{P} \, 2\ran^3 \over \lan 2 \, 3\ran  
\lan 3 \, \hat{P} \ran } 
\right)^2 \ , 
\eeqa
and $P^2 = (p_1 + p_4)^2$.
Using 
\beq
\lan i \, \hat{P} \ran \, = \, 
{\lan i | P | 1] \over [ \hat{P} \, 1 ] } \ , 
\eeq
we find, after a little algebra,  
\beq
\cM^{(4)} \ = \ 
\frac {\lan 1 2 \ran^6 [14] } {\lan 14\ran \lan 23\ran^2
\lan 34\ran^2}
\ . 
\eeq
The full amplitude is $\cM(1^-,2^-,3^+,4^+) = \cM^{(3)} + 
 \cM^{(4)}$. 
Thus, we conclude that the four point MHV amplitude 
generated by our recursion relation is given by
\beq
\cM(1^-,2^-,3^+,4^+)\  = \ 
\frac {\lan 1 2 \ran^6 [14] } {\lan 14\ran \lan 23\ran^2
\lan 34\ran^2}
\, + \, 3 \leftrightarrow 4 
\ . 
\eeq
It is easy to check that this agrees 
with the conventional formula for this amplitude
\beq
\cM(1^-,2^-,3^+,4^+) \ = \ 
\frac {\lan 1 2 \ran^8 [12] } {N(4) \lan 34 \ran}
\ ,
\eeq
where 
\beq
N(n) \ := \  
\prod_{1\leq i < j \leq n} \lan i\, j\ran 
\ , 
\eeq
or, equivalently, with the expression derived from 
the appropriate KLT relation, Eq.~\eqref{4}. 
For the five graviton scattering case, 
our recursion relation yields a sum of three diagrams. 
A calculation similar to that illustrated 
previously for the four-point case leads to the result  
\beq
\label{5point}
\cM(1^-,2^-,3^+,4^+,5^+) \ = \ 
\frac {\lan 12\ran^6 [15][34] } {\lan 15\ran \lan 23\ran
\lan 24\ran\lan 34\ran\lan 35\ran\lan 45\ran} \, +\,  
\cP^{\rm c} (3,4,5)
\ ,
\eeq
where $\cP^{\rm c} (3,4,5)$ means that we 
have to sum over cyclic permutations of the labels $3,4,5$.
The conventional formula for the five graviton 
MHV scattering amplitude is
\beq
\cM(1^-,2^-,3^+,4^+ , 5^+) = \frac {\lan 1 2 \ran^8  } {N(5)}
\Big( [12][34]  
\lan 13\ran\lan 24 \ran-[13][24] \lan 12 \ran\lan 34 \ran\Big)
\ .
\eeq
Using standard spinor identities and momentum conservation,
it is straightforward to check
that our expression \eqref{5point} agrees with this
(alternatively, one can use the KLT relation 
\eqref{5}).

For the six graviton scattering amplitude, 
our recursion relation yields a sum of four terms, 
\beqa
\label{6point}
\cM(1^-,2^-,3^+,4^+,5^+,6^+) & = &  
\frac {\lan 12\ran^6 [16] }{\lan 16\ran}
\cdot {1 \over \lan 2 \, 6\ran 
\lan 3 \, 4\ran \lan 3 \, 5\ran \lan 4 \, 5\ran }
\\ \nonumber 
&&
\hspace{-2.5cm}
\bigg( 
{[3\, 4] \over \lan 2 \, 3\ran  \lan 2 \, 4\ran}
\, 
\frac{\lan 2 | 3+4 |5]}{\lan 56\ran }
 + 
{[4\, 5] \over \lan 2 \, 4\ran  \lan 2 \, 5\ran}
 \, 
\frac{\lan 2 | 4+ 5 |3]}{\lan 36\ran }
 + 
{[5\, 3] \over \lan 2 \, 3\ran  \lan 2 \, 5\ran}
 \, 
\frac{\lan 2 | 5 + 3 |4]}{\lan 46\ran }
\bigg)
\nonumber \\
&& 
\hspace{-2.55cm}  +  \ \cP^{\rm c} (3,4,5,6)
\, .
\nonumber
\eeqa
The known formula for this amplitude is
\beq
\cM(1^-,2^-,3^+,4^+,5^+,6^+) = \lan 12\ran^8 \bigg(
      \frac { [12][45][3|4+5|6\ran  }{ \lan 15\ran
\lan 16\ran\lan 12\ran\lan 23\ran\lan 26\ran\lan 34\ran\lan 36\ran
\lan 45\ran\lan 46\ran\lan 56\ran}
                     + \cP(2,3,4) \bigg)
\ , 
\eeq
where $\cP (2,3,4)$ indicates permutations of the labels 
$2,3,4$.
We have checked  numerically that the  formula \eqref{6point}
agrees with this expression.

%\subsection{Five graviton scattering}

\subsection{General formula for MHV scattering}
Recursion relations of the form given in \cite{bcf}, 
or the graviton recursion relation given here, 
naturally produce general formulae for
scattering amplitudes. 
For a suitable choice of reference spinors, these
new formulae can often be simpler than previously known examples.
For the choice of reference spinors $1,2,$ 
which we have made above,
the graviton recursion relation 
is particularly simple, as it produces only
one term at each step. 
This immediately suggests that one can use it
to generate an explicit expression for the 
$n$-point amplitude.
This turns out to be the case, and
experience with the use of our recursion relation 
leads us to propose the following 
new general formula for the $n$-graviton 
MHV scattering amplitude.
This is 
(labels $1,2$ carry negative helicity, 
the remainder carry positive helicity)
\beq
\label{npoint}
\cM(1, 2, {i_1}, \cdots , {i_{n-2}}) =
\frac{\lan 1\, 2 \ran^6 [1\,  i_{n-2}]}{\lan 1 \, i_{n-2}\ran}\, \, 
G(i_1,i_2,i_3)\,
\prod_{s=3}^{n-3} \frac{\lan 2 | i_1+...+i_{s-1} | i_s]}
{\lan i_s i_{s+1}\ran \lan 2 i_{s+1}\ran }
+ \cP(i_1,...,i_{n-2}),
\eeq
where
\beq
G(i_1,i_2,i_3) = 
{1\over 2} 
\frac{[i_1 i_2]}{\lan 2i_1\ran\lan 2i_2\ran
\lan i_1i_2\ran\lan i_2i_3\ran\lan i_1i_3\ran}
\ .
\eeq
(For $n=5$ the product term is dropped from \eqref{npoint}).
It is straightforward to check that this amplitude satisfies 
the recursion relation with the choice of reference 
legs $1^-$ and $2^-$. 

The known general MHV amplitude for 
two negative helicity gravitons, 1 and 2, and 
the remaining $n-2$ with positive helicity, 
is given by \cite{bgk}
\beq
\cM (1, 2, 3, \cdots , n) =
\lan 12\ran^8 \Biggl[ {[12] [n-2 ~n-1] \over \lan 1~n-1\ran}
{1\over N(n)} \prod_{i=1}^{n-3}
\prod_{j=i+2}^{n-1} \lan ij\ran ~F
\ +\ {\cal P} (2,\ldots ,n-2)\Biggr],
\label{15}
\eeq
where
\beq
F= \left\{ \begin{array}{ll}
\prod_{l=3}^{n-3}\,  [l| (p_{l+1} + p_{l+2}+\cdots
+p_{n-1}) | n\ran &
n\geq 6\\
1& n=5\\
\end{array}
\right.\label{16}
\eeq
We have checked numerically, up to $n=11$, 
that our formula \eqref{npoint}
gives the same results as \eqref{15}.

\section{Applications to other field theories}
One of the striking features of the BCFW proof 
of the BCF recursion relations is that it is almost
not needed to specify the theory with which we are dealing.
Indeed, in \cite{bcfw} the only step where 
specifying the theory did matter was
in the estimate of the behaviour of 
the scattering amplitudes $\cM (z)$ as $z \to \infty$, 
which was important to assess the possible 
existence of boundary terms in the recursion relation.
This leads us to conjecture that recursion relations 
could be a more generic feature of massless 
(or spontaneously broken) 
field theories in four dimensions.%
\footnote{This was also suggested in \cite{Bern:2005hs}.} 
After all, the BCF recursion relations, as well as the 
recursion relation for gravity amplitudes 
discussed in this paper, just reconstruct 
a tree-level amplitude (which is a rational function) 
from its poles. 

Let us focus on massless $\l (\phi^\dagger \phi)^2$ 
theory in four dimensions. 
We use the spinor helicity formalism, meaning that 
each momentum will be written as 
$p_{a \dot{a}} = \l_a \lt_{\dot{a}}$.
A scalar propagator $1/P^2$ connects states of 
opposite ``helicity'', 
which here just means that the propagator is 
$\lan \phi (x) \phi^\dagger (0) \ran$, with 
 $\lan \phi (x) \phi (0) \ran = 
\lan \phi^\dagger (x) \phi^\dagger  (0) \ran = 0$.
Now consider a Feynman diagram contributing to an 
$n$-particle scattering amplitude, 
and let us shift the momenta of 
particles $k$ and $l$ as in \eqref{shifts}. 
As for the Yang-Mills case discussed in 
\cite{bcfw}, there is a unique path of propagators 
going from particle $k$ to particle $l$.  
Each of these propagators contributes $1/z$ at large $z$, 
whereas vertices are independent of $z$. 
We thus expect Feynman diagrams contributing to 
the amplitude to vanish in the large-$z$ limit. 

An exception to the above reasoning is represented 
by those Feynman diagrams where the shifted legs belong 
to the same vertex; these diagrams are $z$-independent, 
and hence not suppressed as $z \to \infty$. 
In order to deal with this problematic situation, and ensure that 
the full amplitude computed from Feynman diagrams $\cM (z)$ 
vanishes  as $z \to \infty$ we propose two alternatives.
 
Firstly, if one considers  $(\phi \phi^\dagger)^2$
theory without any group structure, 
one can remove the problem by performing multiple shifts. 
This possibility has already been used in the  context 
of the rational part of one-loop amplitudes in pure Yang-Mills
\cite{Bern:2005hs}. 
In our case, it is sufficient to shift at least four 
external momenta. 
% one has to shift at least four external moment. 

Alternatively, we can consider $(\phi \phi^\dagger)^2$ theory 
with global symmetry group $U(N)$ and $\phi$ in the adjoint. 
In this case we can group the amplitude into
colour-ordered partial amplitudes, as in the Yang-Mills case. 
Then, for any colour-ordered amplitude one can always find 
a choice of shifts such that the shifted legs do not belong  
to the same Feynman vertex. 
% for this particular subclass of
%colour-ordered Feynman diagrams. 
The procedure can be repeated for any colour ordering, 
and the complete amplitude is obtained
by summing over non-cyclic permutations of the external legs.
% this solving the problem. 
%For example,  for the ordering $1,2,3,4,5,6, \ldots$,  
%we could choose legs $1$ and $4$ as reference legs; 
%then there are no Feynman diagrams where
%these legs are part of the same vertex, and 
%the partial amplitude $M(z)$ vanishes as $z \to \infty$. 

In this way, the appearance of a boundary term $\cC_{\infty}$ 
can be avoided, and 
one can thus derive a recursion relation for 
scattering amplitudes similar to \eqref{rec}.
A similar analysis can be carried out in other theories, 
possibly in the presence of spontaneous symmetry breaking, etc. 
We expect this to play an important r\^{o}le in future studies. 

%%%%%%%%%%%%%%%%%%%%%%%%%%%%%%%%%%%%%%%%%%%%%%%%%%%%%%%%%%%

\section*{Acknowledgements}

It is a pleasure to thank Valya Khoze, Marco Matone and
Sanjaye Ramgoolam for discussions.
GT acknowledges the support of PPARC.

%\newpage

%%%%%%%%%%%%%%%%%%%%%%%%%%%%%%%%%%%%%%%%%%%%%%%%%%%%%%%%%
\startappendix
\Appendix{KLT relations}
For completeness, in this appendix 
we write  the field theory limit of the KLT relations
\cite{klt} for the case of four, five and six points: 
\beqa
\label{3}
\cM (1,2,3) & = & 
-i\cA (1,2,3)\, \cA (1,2,3)
\ , 
\\  \cr
\cM (1,2,3,4) & = & 
-is_{12}\ \cA (1,2,3,4)\cA(1,2,4,3) \ , 
\label{4}
\eeqa
\beqa
\label{5}
\cM (1,2,3,4,5)  & = & 
is_{12}s_{34}\ \cA (1,2,3,4,5)\cA(2,1,4,3,5) 
\nonumber \\
&+&i s_{13}s_{24}\ \cA(1,3,2,4,5)\cA(3,1,4,2,5) \ , 
\\ \cr  
\label{6}
\cM (1,2,3,4,5,6)& = &
-is_{12}s_{45}\ \cA(1,2,3,4,5,6)
\big[ s_{35}\cA(2,1,5,3,4,6)
\nonumber \\
 & + & (s_{34}+s_{35})\ \cA (2,1,5,4,3,6)\big] 
\\ \nonumber &+ & \cP (2,3,4) \ . 
\eeqa
In these formulae, $\cM$ ($\cA$) denotes 
a tree-level gravity (Yang-Mills colour-ordered) 
amplitude, $s_{ij} : = (p_i + p_j)^2$, 
and $\cP (2,3,4)$ stands for permutations of $(2,3,4)$.  
The relation for a generic number of particles can be found in 
 \cite{Bern:1998sv}.

%%%%%%%%%%%%%%%%%%%%%%%%%%%%%%%%%%%%%%%%%%%%%%%%%%%%%%%%%

\end{document}